\documentclass[fleqn,usenatbib]{mnras}

\usepackage{newtxtext,newtxmath}

\usepackage[T1]{fontenc}
\usepackage{ae,aecompl}
\newcommand{\bc}{}

\usepackage{graphicx}	
\usepackage{amsmath}	
\usepackage{amssymb}	

\title[Young planets]{Constraining the entropy of formation from young transiting planets}

\author[Owen, J. E]{
James E. Owen\thanks{E-mail: james.owen@imperial.ac.uk}
\\
Astrophysics Group, Imperial College London, Blackett Laboratory, Prince Consort Road, London SW7 2AZ, UK\\
}

\date{Accepted XXX. Received YYY; in original form ZZZ}

\pubyear{2015}

\begin{document}
\label{firstpage}
\pagerange{\pageref{firstpage}--\pageref{lastpage}}
\maketitle

\begin{abstract}
Recently {\it K2} and {\it TESS} have discovered transiting planets with radii between $\sim$ 5-10~R$_\oplus$ around stars with ages $<100$~Myr. These young planets are likely to be the progenitors of the ubiquitous super-earths/sub-neptunes, that are well studied around stars with ages $\gtrsim 1$~Gyr. The formation and early evolution of super-earths/sub-neptunes are poorly understood. Various planetary origin scenarios predict a wide range of possible formation entropies. We show how the formation entropies of young ($\sim$20-60~Myr), highly irradiated planets can be constrained if their mass, radius and age are measured.  This method works by determining how low-mass a H/He envelope a planet can retain against mass-loss. This lower bound on the H/He envelope mass can then be converted into an upper bound on the entropy. If planet mass measurements with errors $\lesssim 20$\% can be achieved for the discovered young planets around DS Tuc A and V1298 Tau, then insights into their origins can be obtained. For these planets, higher measured planet masses would be consistent with standard core-accretion theory. In contrast, lower planet masses ($\lesssim 6-7$~M$_\oplus$) would require a ``boil-off'' phase during protoplanetary disc dispersal to explain. 
\end{abstract}




\section{Introduction}

Exoplanet surveys and demographic studies have revealed a previously unknown class of planets. These close-in, super-earth/sub-neptunes have radii in the range 1-4~R$_\oplus$ \citep[e.g.][]{Bourucki2011,Thompson2018} and masses $\lesssim 20~$M$_\oplus$ \citep[e.g.][]{Mayor2011,Wu2013,Marcy2014,Weiss2014,Hadden2014,JontofHutter2016}. With orbital periods of less than a few hundred days, these planets have been shown to be incredibly common, with most sun-like and later-type stars hosting at least one, if not many \citep[e.g.][]{Fressin2013,Silburt2015,Mulders2018,Zink2019,Hsu2019}. 

Combined mass and radius measurements for individual planets revealed a vast spread in densities; from as high as $\sim 10$~g~cm$^{-3}$ to as low as $\sim 0.1~$g~cm$^{-3}$. The former is consistent with rocky bodies with Earth-like (iron/silicate) compositions \citep[e.g.][]{Dressing2015,Dorn2019} and the latter with solid cores surrounded by larger H/He atmospheres \citep[e.g.][]{JontofHutter2016}. At intermediate densities a plethora of compositions are possible, including some combination of iron, silicates, water and H/He \citep[e.g.][]{Rogers2010,Zeng2019}. 

H/He envelopes on low-mass planets close to their host stars are vulnerable to mass-loss \citep[e.g.][]{Lammer2003,Baraffe2005,MurrayClay2009,owen12}. Evolutionary models of solid-cores surrounded by H/He envelopes suggested that mass-loss would sculpt the planet population into two classes: those which completely lose their H/He envelope leaving behind a ``stripped core'', and those planets that retain about 1~\% by mass in their H/He envelope \citep[e.g.][]{Owen2013,Lopez2013,Jin2014}.  Early indications of such a dichotomy were found in planetary density measurements \citep[e.g.][]{Weiss2014,Rogers2015}. However, it was not until accurately measured stellar properties allowed precise planetary radii to be determined that a bi-modal radius distribution was revealed \citep[e.g.][]{Fulton2017,Fulton2018,VanEylen2018}. 

Incorporating these evolutionary clues into the compositional determination suggests most planets larger than about 1.8~R$_\oplus$ consist of an Earth-like core surrounded by a H/He envelope, where this envelope contains a few percent of the planet's mass \citep{Wolfgang2015}. Further, mass-loss models suggest that the vast majority of those planets that do not possess H/He envelopes today, formed with one which they then lost \citep[e.g.][]{Owen2017,Wu2019,Rogers2020}. 

The majority of exoplanet host stars are billions of years old. Recent work by \citet{Berger2020} has shown that the median age of the {\it Kepler} planet host stars is $\sim 4$~Gyr and only $\sim 3$\%
of the {\it Kepler} host stars are $<1$~Gyr old. Therefore, most of our knowledge about the demographics of close-in exoplanets is restricted to old stars. The fact many of these planets possess H/He envelopes necessitates they formed before gas-disc dispersal. These gas-discs have lifetimes of $1-10$~Myr \citep[e.g.][]{haisch01,Mamajek2009}. {\bc Thus, most exoplanet host-stars are significantly older than the planetary formation timescale ($\lesssim 10$~Myr). }

Compositional uncertainty and lack of knowledge of the exoplanet demographics has led to considerable discussion and debate about how these planets form. In many cases the size of the H/He envelopes implies they must have accreted from a protoplanetary disc. The idea of accretion of a H/He envelope by a solid core fits within the general framework of core-accretion \citep[e.g.][]{Rafikov2006,Ikoma2006}. In the core-accretion model, once the planet is massive enough that its Bondi-radius resides outside its physical radius (crudely at a few tenths of a Lunar mass, e.g. \citealt{Massol2016}) it can gravitationally bind nebula gas. As the planet's solid mass continues to grow it can bind ever larger quantities of gas; {\bc eventually at around core masses of $\sim$0.5~M$_\oplus$ the H/He envelope's cooling time becomes comparable to the disc's lifetime \citep[e.g.][]{Lee2015}.} Thus, beyond solid core masses of $\sim$ 0.5~M$_\oplus$\footnote{Obviously this precise value is uncertain and depends on details of the opacity, e.g. \citealt{Venturini2016,LeeConnors2020}.} accretion of a H/He envelope is dependant on how fast current gas in the planet's envelope evolves thermally; heat the envelope gas via solid accretion, and gas flows from the envelope to the disc, let it cool and contract, and high-entropy nebula gas flows into the envelope. 

The expected thermal envelope structure typically consists of a convective zone surrounding the core, where heat generated by solid accretion or gravitational contraction of the atmosphere is rapidly convected outwards. {\bc Eventually, as the temperature and density drops a radiative zone forms, which connects the envelope to the disc}, where the lower entropy interior is connected to the higher entropy disc \citep[e.g.][]{Rafikov2006}. Since opacity typically increases with pressure and temperature it's the radiative-convective boundary that controls the thermal evolution of the envelope, and hence the rate of accretion. This makes gas accretion fairly insensitive to the nebula's properties (e.g. density and temperature). 

Despite the basic success of the core-accretion model in explaining that low-mass planets can acquire a H/He envelope, it has failed to quantitatively explain the properties of observed sub-Neptunes \citep[e.g.][]{Lee2014,Ogihara2020,Alessi2020}. In fact the issue is not that they have accreted H/He atmospheres, it is that they have only accreted a few percent by mass, even after mass-loss processes like photoevaporation \citep{Jankovic2019,Ogihara2020} and/or core-powered mass-loss \citep[e.g.][]{Ginzburg2018} have been taken into account. Specifically, \citet{Rogers2020} has recently shown standard core-accretion models over-predict the H/He envelope mass by a factor of $\sim$5 at the end of disc dispersal.

These problems have led to several proposed solutions: either slowing the accretion of H/He \citep[e.g.][]{Ormel2015,Lee2016,Ginzburg2017,Chen2020}, typically by increasing the entropy of the interior, or including extra, rapid mass-loss, most notably {\it during} protoplanetary disc dispersal, decreasing the entropy of the interior \citep[e.g.][]{OW2016,Ginzburg2016,Fossati2017,Kubyshkina2018_young} (see the discussion for a detailed description of these models).    More fundamentally, these proposed solutions all modify the thermodynamic evolution of the H/He envelope. Without any intervention one would expect the planet to have an initial cooling timescale (or Kelvin-Helmholtz contraction timescale) comparable to the time over which it has been forming, i.e. a few million years. Thus, in the first solution where accretion is slowed by increasing the entropy of the interior, the envelope's cooling time becomes shorter (probably less than a few million years). In the second solution, where rapid mass-loss during disc dispersal decreases the {\bc planet's radius and entropy}, planets end up with much longer cooling times, closer to $\sim 100$~Myr \citep{OW2016}. This mechanism is known as ``boil-off''.  

As planets age they cool; by the time they reach a Gyr old they gone through many {\bc initial} cooling times and have completely forgotten their initial thermodynamic properties. Therefore, our population of old planets is generally unable to tell us about the thermodynamic properties of forming or recently formed planets. 

With the advent of wider searches for transiting exoplanets (e.g. K2 \citealt{K2Mission}, TESS \citealt{TESSmission} \& NGTS \citealt{NGTS}) discovering young ($\lesssim 100$~Myr) close-in planets has become possible. Notable recent examples are the four planets in the V1298 Tau system at an age of $\sim$ 24~Myr \citep{David2019}, a $\sim 45$~Myr old planet around DS Tuc A \citep{Newton2019}, a $\sim 23$~Myr old planet around AU Mic \citep{Plavchan2020} as well as other recent results from the {\sc thyme} project \citep[e.g.][]{THYMEII}.

In this work, we show how the combination of a mass and radius measurement for young planets can be used to place a constraint on their initial entropy. In Section~2 we demonstrate how young planets with measured mass and radii can be used to constrain their initial entropies, using simple models. In Section~3 we use numerical planetary evolution models to explore this further, investigating real systems in Section~4. In Section~5 we discuss the implications of our work and summarise in Section~6.

\section{A sketch of the idea}
\begin{figure}
    \centering
    \includegraphics[width=\columnwidth]{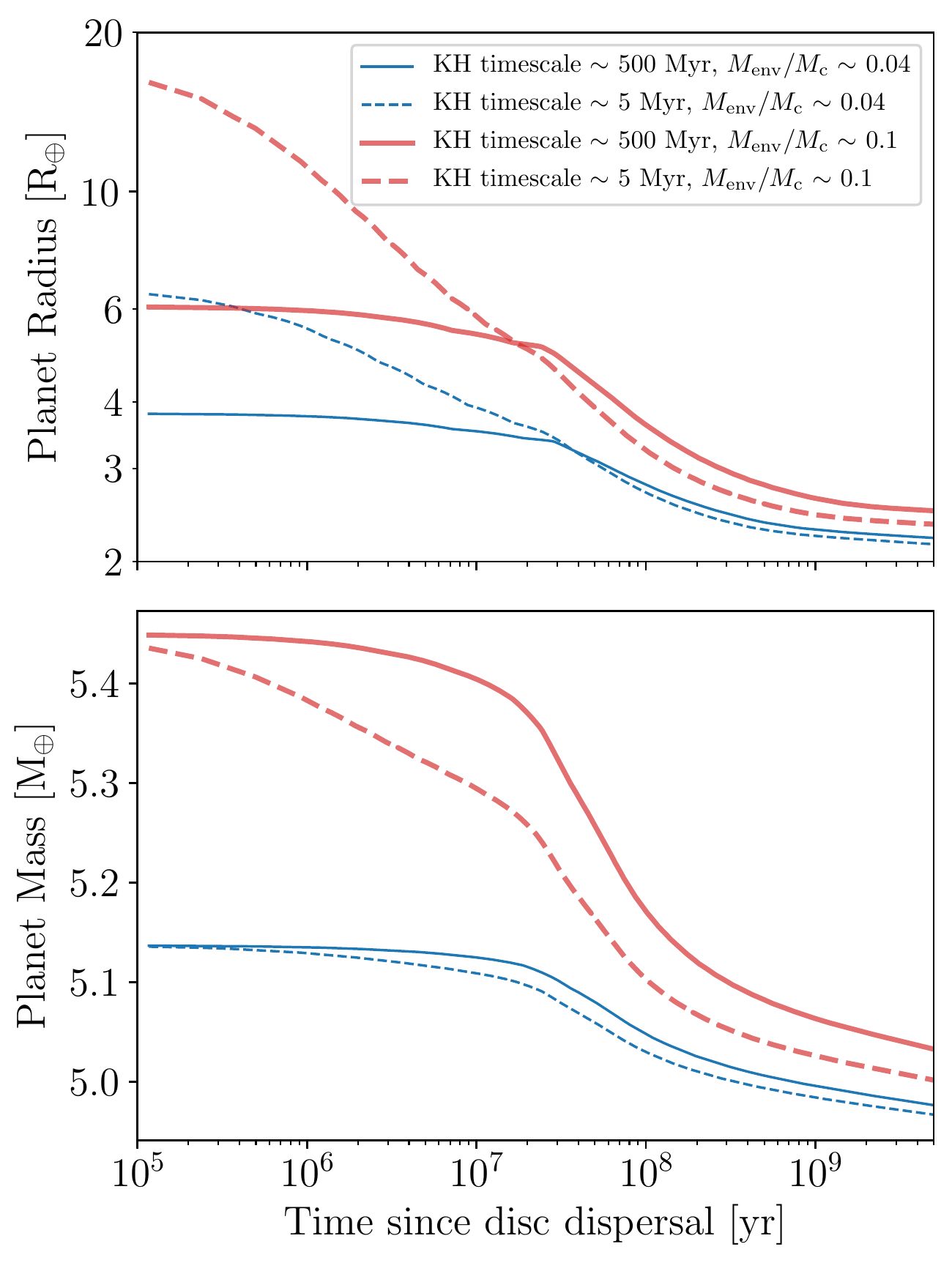}
    \caption{{\bc The planet's radius (top) and total planet mass (bottom)} evolution of low-mass, close-in exoplanets with H/He atmospheres. The solid and dashed lines show planets whose envelopes have initial Kelvin-Helmholtz contraction timescales of 500 and 5 Myr respectively. The thick and thin lines show planets that have initial envelope fractions ($M_{\rm env}/M_{\rm c}$) of 0.1 and 0.04 respectively. The evolution calculations are computed using {\sc mesa} as described in Section~\ref{sec:mesa} and include thermal evolution and photoevaporation. The planet contains a core with a mass of $\sim 5$~M$_\oplus$ and is located at 0.1~AU around a sun-like star.}
    \label{fig:basic_evol}
\end{figure}

Present day sub-neptunes with voluminous H/He atmospheres cool over time, contracting under the release of heat left over from formation (both in their atmospheres and solid cores).
Evolutionary tracks \citep[e.g.][]{Baraffe2005,Lopez2014} predict these planets were substantially larger when younger, even when mass-loss is not factored in. In Figure~\ref{fig:basic_evol} we show tracks from planetary evolution models for planets that evolve into typical sub-neptunes after billions of years of evolution. These planets have radii in the range 4-15~R$_\oplus$ at ages $\lesssim 100$~Myr.  

In fact the upper envelope of the radius evolution for present-day sub-neptunes is expected to become $\gtrsim 1~$R$_{\rm Jup}$ at young ages. This is demonstrated in Figure~\ref{fig:basic_evol} indicating that at the youngest ages these planets could be conflated with a giant planet, as they have radii similar to present day hot jupiters. In fact, young hot jupiters are expected to have radii in excess of $\sim$ 13~R$_\oplus$ \citep{Fortney2010}. This indicates that even those young planets that have measured radii $\sim 10$~R$_\oplus$ (e.g. V1298 Tau b, \citealt{David2019} and HIP 67522b, \citealt{THYMEII}) are likely to be ``proto-sub-neputunes'' rather than young, jupiter mass planets.  

Unlike stars, planet's have no method of preventing gravitational collapse (other than Coulomb or degeneracy pressure), therefore, the observed size of a planet depends on the thermodynamic state of it's interior, or how much entropy it currently possesses. As the planet's envelope contracts the interior entropy falls. Eventually, the interior will either be supported by degeneracy pressure at high masses or, at low-masses by Coulomb pressure \citep{Seager2007}.

It is well known that with measurements of the mass and radius of a single composition self-gravitating sphere (e.g. a pure H/He mixture), then the internal thermodynamic state and interior can be determined. However, both observational \citep{WL12,JontofHutter2016,Fulton2017,VanEylen2018,Benneke2019,Benneke2019b}, and theoretical \citep{Owen2017,Wu2019,Rogers2020} evidence suggests present-day sub-neptunes are not of a single composition, but mostly likely composed of a solid core surrounded by a H/He envelope\footnote{Although alternative ideas do exists, e.g. \citet{Zeng2019}.}. As we show below, this makes the envelope mass and its internal thermodynamic state degenerate. Specifically, a planet with a known mass and radius can have more mass in its core and less in its envelope if the envelope is hotter (and therefore has a larger scale height) or vice-versa, {\bc as demonstrated by works where extra heating mechanisms are included \citep[e.g.][]{Pu2017,Millholland2020}}. This is even before the degeneracy between the composition of the core and envelope mass is taken into account.   

\subsection{Degeneracy between internal entropy and envelope mass fraction}

The degeneracy between entropy and envelope mass fraction is trivial to identify. Given a planet composed of a core of known density surrounded by a H/He envelope, there are three parameters which specify its structure: core mass, envelope mass and internal entropy. Thus, with only measurements of a planet's mass and radius it is not possible to constrain these three structure parameters, and thus we cannot determine its internal entropy.  

This can be seen explicilty if we build a simple model for the planet's internal structure. This simple model is based on the assumption that self-gravity can be neglected in the envelope\footnote{It can also be seen by considering ``loaded'' polytropes \citep{Huntley1975} which include atmosphere self-gravity and smoothly tend to the mono-composition models as the envelope mass exceeds the core mass. However, such analysis is not necessary to elucidate the basic point.}\citep[e.g.][]{Rafikov2006,Piso2014,Lee2015,Ginzburg2016}. As derived by \citet{Owen2017} \& \citet{OCE2020} the envelope mass, $M_{\rm env}$, surrounding a core of mass $M_c$ and radius $R_c$ with equation-of-state relating pressure, $P$ to density, $\rho$ via $P\propto \rho^\gamma$ is given, approximately by:
\begin{equation}
    M_{\rm env}\approx 4\pi R_{\rm rcb}^3\rho_{\rm rcb}\left(\nabla_{\rm ab}\frac{GM_c}{c_s^2 R_{\rm rcb}}\right)^{1/(\gamma-1)} I_2\left(R_c/R_{\rm rcb},\gamma\right) \label{eqn:Menv}
\end{equation}
where $R_{\rm rcb}$ is the radius of the radiative-convective boundary, $\nabla_{\rm ab}$ is the adiabatic gradient, $c_s$ the isothermal sound-speed at the radiative-convective boundary and $I_n$ is a dimensionless integral of the form:
\begin{equation}
    I_n\left(R_c/R_{\rm rcb},\gamma\right)=\int_{R_c/R_{\rm rcb}}^1 x^n\left(x^{-1}-1\right)^{1/(\gamma-1)}{\rm d}x
\end{equation}
Due to the extreme irradiation level most close-in planets receive, the radiative-convective boundary occurs at optical depths much higher than the photosphere. The radiative-convective boundary sets the point at which the energy released by gravitational contraction in the interior is no-longer transported by the convection, but rather by radiation. Hence the luminosity of the planet (and therefore internal entropy) is related directly to the density at the radiative-convective boundary. Since luminosity and entropy are rather non-intuitive quantities and do not facilitate easy comparison across planet-mass, age or formation models we follow \citet{Owen2017} and choose the atmosphere's Kelvin-Helmholtz contraction timescale $\tau_{\rm KH}$ (or cooling timescale) as our quantity to describe the entropy and thermodynamic state of the planetary interior. We choose this parameterisation of entropy as this cooling timescale can be directly compared to quantities like the protoplanetary disc lifetime. This allows us to write the luminosity as:
\begin{equation}
    L\approx\frac{1}{\tau_{\rm KH}}\frac{GM_cM_{\rm env}}{R_{\rm rcb}}\frac{I_1}{I_2}
\end{equation}

For clarity, we restrict ourselves to a constant opacity envelope, with opacity $\kappa$ (while this is clearly incorrect, nothing we demonstrate below is invalidated by this). Thus, solving for the density at the radiative-convective boundary \citep[e.g.][]{Owen2017} we find:
\begin{equation}
    \rho_{\rm rcb}\approx\left(\frac{\mu}{k_b}\right)\left[\left(\frac{I_2}{I_1}\right)\frac{64\pi\sigma T^3 R_{\rm rcb} \tau_{\rm KH}}{3\kappa M_{\rm env}}\right] \label{eqn:rho_rcb}
\end{equation}
with $\mu$ the mean molecular weight, $k_b$ Boltzmann's constant, $\sigma$ Stefan-Boltzmann's constant and $T$ is the equilibrium temperature of the planet.  Finally, noting that the radiative layer is approximately isothermal (and therefore, well described by an exoponential density profile) and thus thin, we simply take $R_{\rm rcb}\approx R_p$. Combining Equations~\ref{eqn:Menv} and \ref{eqn:rho_rcb} we can derive a mass-radius relationship for our planets of the form:

\begin{equation}
    R_p^{(4\gamma - 5)/(\gamma -1)}M_p^{1/(\gamma -1)}\propto M_{\rm env}^2 \tau_{\rm KH}^{-1}\frac{I_1}{I_2^2} \label{eqn:mass_radius}
\end{equation}
where we have dropped unimportant variables. Equation~\ref{eqn:mass_radius} clearly demonstrates that even with the measurement of a planet's mass and radius, the Kelvin-Helmholtz contraction timescale cannot be determined. Now, in the limit $M_{\rm env}\rightarrow M_p$ this degeneracy disappears\footnote{However, one needs to include self-gravity in the analysis.}. For old planets this degeneracy is bypassed by the reasonable assumption that the planet has cooled sufficiently that its initial thermodynamic state has been forgotten and $\tau_{\rm KH}$ has simply tended towards the age of the planet ($T_{\rm age}$). However, for young planets such statements cannot be made, all we can safely say is $\tau_{\rm KH}\gtrsim T_{\rm age}$. Thus, at young ages, measurements of a planet's mass and radius are not sufficient to determine either its internal composition (fraction of mass in the atmosphere compared to the core) or its internal thermodynamic state.

 The dependence in the core composition is encapsulated in the dimensional integrals $I_1$ and $I_2$. In the limit $R_p/R_c \gg 1$, relevant for young planets, both integrals tend to a constant for $\gamma > 3/2$ and $I_2$ tends to a constant for $>4/3$. Inspection of our numerical models (Section~\ref{sec:mesa}) indicates that planetary interiors span the full range of possible limits, with $\gamma <4/3$ close to the planetary cores when the interiors are high-entropy and $\gamma > 3/2$ closer to the planetary surface for low-entropy interiors. In order to assess whether core-composition will effect our analysis we calculate how the ratio $I_2^2/I_1$ varies with core-composition at different values of $\gamma$, we do this for a 7~R$_\oplus$, 5~M$_\oplus$ planet. For an extreme variation in core composition of 1/3 ice, 2/3 rock to 1/3 iron, 2/3 rock we find a variation of 4\% for $\gamma=5/3$ and a factor of two for a of $\gamma=1.25$ (the lowest found at any point in the numerical models). These variations are much smaller than the order of magnitude changes in the Kelvin-Helmholtz timescale we are investigating, especially when considering detailed fits to the exoplanet data suggest the spread in core-composition is narrow \citep[e.g.][]{Dorn2019,Rogers2020}. Specifically, for the spread in core composition inferred by \citet{Rogers2020} the ratio $I_2^2/I_1$ varies by a maximum of 15\% for $\gamma=1.25$ for a 7~R$_\oplus$, 5~M$_\oplus$ planet. Thus, we consider our results to be robust to uncertainties in the core-composition, and certainly smaller than variations arising from the observational uncertainties on age, mass and radius\footnote{In fact retaining the dimensional integrals and an arbitrary choice of $\gamma$, we find the dependence on the constrain of $\tau_{\rm KH}$ in Equation~\ref{eqn:critera2} scales linearly with $I_2^2/I_1$, compared to much higher powers of mass, radius and age, indicating that for typical 10-20\% errors, the variation in the dimensional integrals with core-compositions will have a small effect on our constraint on the Kelvin-Helmholtz timescale. }. Given we have already adopted a constant opacity, we chose to adopt a constant value of $\gamma=5/3$ and ignore the variation of $I_2$ and $I_1$ for simplicity in the rest of the section, while noting no choice of a single value of $\gamma$ is justifiable. We emphasise that this section is purely an illustrative demonstration of the method and these choices do not affect the general idea. In our numerical models in Section~\ref{sec:mesa} the appropriate equation-of-state and opacities are used.

\subsection{Leveraging mass-loss}

Fortunately, for close-in planets we do have a way of constraining the mass in a planet's envelope. The high-irradiation levels experienced by young planets cause them to lose envelope mass over time \citep[e.g.][]{Baraffe2005,Lopez2013,Owen2013}. Specifically, given a planet with a known mass and radius there is a minimum envelope mass it could have retained given its age. Make the envelope less massive and it could not have survived mass-loss until its current age. Therefore, the envelope mass-loss timescale $t_{\dot{m}}$, must satisfy:
\begin{equation}
t_{\dot{m}}\equiv\frac{M_{\rm env}}{\dot{M}}\gtrsim T_{\rm age}\label{eqn:tmdot1}
\end{equation}
Since the mass-loss rate $\dot{M}$ depends only on planet mass and radius (for externally driven loss processes), or planet mass, radius and Kelvin-Helmholtz contraction timescale (for internally driven loss processes such as core-powered mass-loss \citealt{Ginzburg2018,Gupta2019}), then the inequality in Equation~\ref{eqn:tmdot1} can be used to place a lower bound on the Kelvin-Helmholtz timescale of the planet. Working within the framework where the loss is driven by photoevaporation, we show that with a planet radius alone one can place a minimum value on the planet mass to be consistent with the simplest (``vanilla'') picture of planetary formation via core accretion. Further, with a measured mass and radius we demonstrate how a lower bound on the Kelvin-Helmholtz timescale can be found.  
\subsection{A minimum mass for the ``vanilla'' scenario}  
\label{sec:min_mass}
The most na\"ive expectation is that planet formation is smooth and continuous and disc dispersal gently releases the planet into the circumstellar environment wherin mass-loss can proceed. In this scenario, with no violent processes, the planet's Kelvin-Helmholtz contraction timescale should roughly track age. Therefore, at young-ages we can follow what is done at old ages, where we accept that a planet is several cooling times old and set $\tau_{\rm KH}\sim T_{\rm age}$. Combining this anszat with Equation~\ref{eqn:mass_radius} and inequality \ref{eqn:tmdot1} we find:

\begin{equation}
    M_p^{3/4}\gtrsim A\, \dot{M}T_{\rm age}^{1/2}R_p^{-5/4}
\end{equation}
where the term $A$ incorporates all the terms we have dropped (e.g. temperature, opacity, mean-molecular weight and fundamental constants). Setting the mass-loss rate to:
\begin{equation}
   \dot{M}\propto \frac{F_{\rm HE}\pi R_p^3}{G M_p}\propto R_p^3/M_p
\end{equation} as found in the case of energy-limited photoevaporation ({\bc with $F_{\rm HE}$ the high energy flux received by the planet),} we find:
\begin{equation}
M_p \gtrsim A'\, R_p T_{\rm age}^{2/7} \label{eqn:criterion}
\end{equation}
with $A'\equiv A^{4/7}$. Put simply, for a young planet with a measured radius from a transit survey there is a minimum mass for it to be consistent with the na\"ivest expectation of planet formation. Put another way, if a planet is consistent with the criterion in Equation~\ref{eqn:criterion}, then limited constraints can be placed on its formation entropy and history. Noting in the scenario above where $T\propto a^{-1/2}$ (with $a$ the orbital separation) and $\dot{M}\propto a^{-2}$ we find $A' \propto a ^{-13/7}$, validating the expectation that the higher irradiation levels closer to the star lead to higher mass-loss and therefore higher required planet masses. Finally (in the case of photoevaporation), the rapid drop in XUV flux when the star spins down \citep[e.g.][]{Tu2015} means $T_{\rm age}$ cannot be set arbitrarily long, but is rather constrained to be the saturation time of the XUV output of the star. Hence, the typically quoted values of $\sim 100$~Myr for sun-like stars \citep[e.g.][]{Jackson2012}. 

While the above style of calculation is unlikely to provide interesting analysis for real planets, it could be useful for selecting which planets to target with radial velocity, transit-timing variation (TTV) or spectroscopic follow-up.


\subsection{Constraining entropy of formation}
\label{sec:min_tkh}
Doing away with the anszat that $\tau_{\rm KH}\sim T_{\rm age}$, we can now generalise to the possibility that at young ages $\tau_{\rm KH}\gtrsim T_{\rm age}$. Now we can follow a similar argument to that in the preceeding section, and show that with measurement of a planet's mass and radius, one can place a lower bound on the planet's Kelvin-Helmholtz contraction timescale. 

Again combining Equation~\ref{eqn:mass_radius} for the mass-radius relationship  with the mass-loss criterion in Equation~\ref{eqn:tmdot1} we find:
\begin{equation}
    \tau_{\rm KH} \gtrsim B\, R_p^{7/2} M_p^{-7/2} T_{\rm age}^2 \label{eqn:critera2}
\end{equation}
where like the $A$ factor above, $B$ encapsulates all the terms and fundamental constants we have dropped from our analysis. The dependence of the inequality in Equation~\ref{eqn:critera2} is easy to understand. Larger and less massive planets which are older have experienced more mass-loss. The higher total mass-loss requires a higher atmosphere mass to resist, necessitating a lower entropy interior to give a planet with the same total mass and radius (Equation~\ref{eqn:mass_radius}). Again for the case where $T\propto a^{-1/2}$ and $\dot{M}\propto a^{-2}$ we find $B \propto a ^{-13/4}$ indicating that it is those planets that are closest to their host stars (and experience more vigorous mass-loss) that are the most constraining. Now clearly, if one finds a constraint on the Kelvin-Helmholtz timescale that is shorter than its age, one has not learnt anything other than it is consistent with the ``vanilla'' scenario for core-accretion, and satisfies the constraint in Equation~\ref{eqn:criterion}.  

With a sample of young planets with ages less than a few 100~Myr with measured masses and radii it is possible to constrain their current Kelvin-Helmholtz contraction timescales and hence gain insights into their formation entropies and the physical processes that lead to their formation and early evolution. On the flip-side, if {\it all} young planets appear to be consistent with $\tau_{\rm KH}\sim T_{\rm age}$ at young ages we can also make inferences about their formation pathways.

We caution that in the previous sections we deliberately chose an incorrect opacity-law (a constant opacity) and a simple mass-loss model, in order that the powers in the previous expressions did not become large integer ratios, and as such they should not be used for any quantitative analysis. Switching to more realistic opacity and mass-loss laws does not change the facts identified in Section~\ref{sec:min_mass} and \ref{sec:min_tkh}. 

\subsection{A slightly more sophisticated demonstration}

Before we switch to using full numerical solutions of planetary evolution we can get a sense of the range of interesting planet properties by using the semi-analytic planet structure model developed by \citet{Owen2017}, where all choices (opacity-law, mass-loss model etc) follow those in \citet{OCE2020}. In all cases we assume an Earth-like core composition with a 1/3 iron to 2/3 rock mass-ratio, which is consistent with the current exoplanet demographics (but as mentioned above, such a choice does not strongly affect our results).  
\subsubsection{Minimum masses}
\begin{figure}
    \centering
    \includegraphics[width=\columnwidth]{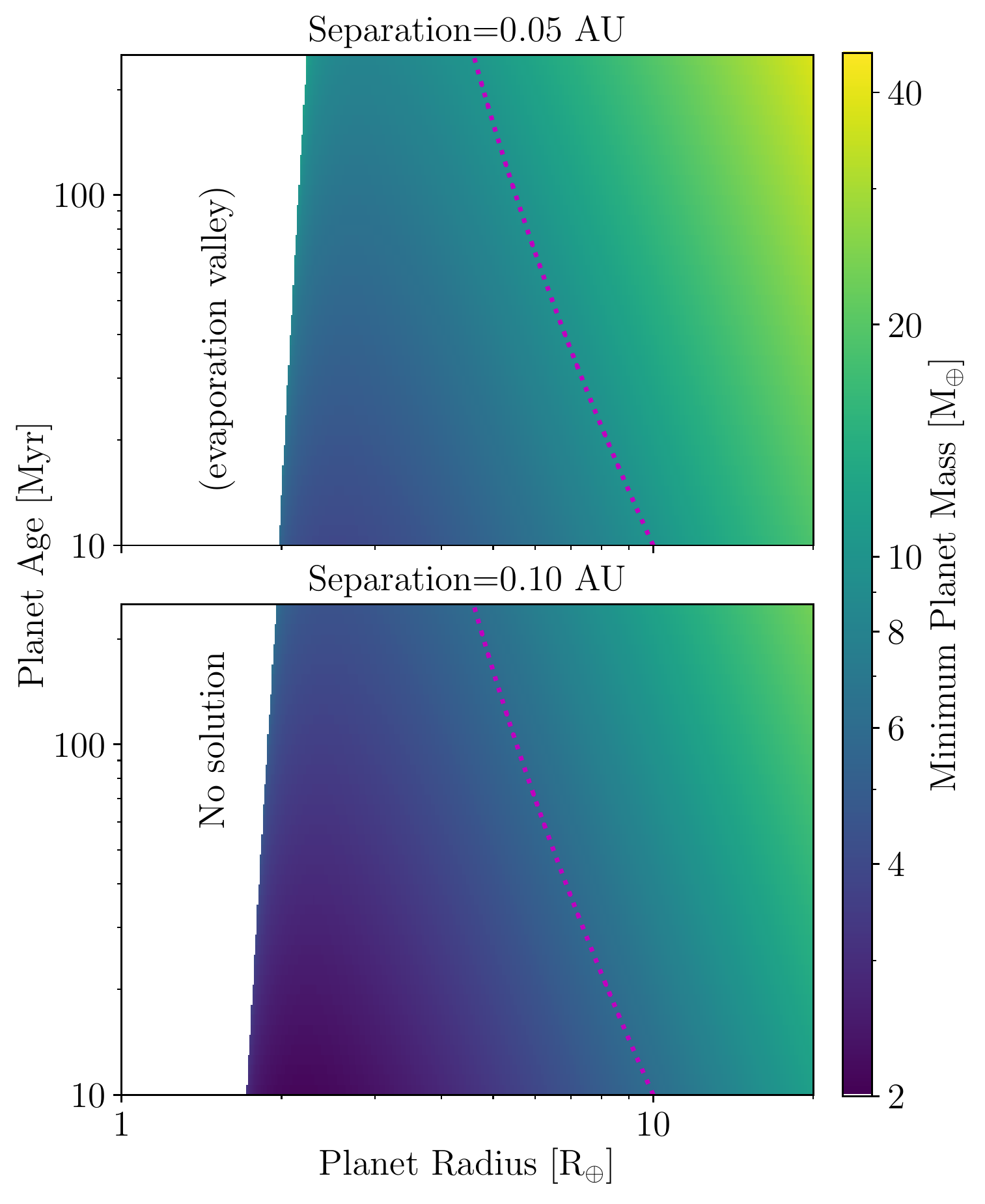}
    \caption{The minimum mass required to be consistent with a scenario where $\tau_{\rm KH}\sim T_{\rm age}$ at all ages.{\bc This figure uses the more sophisticated calculation described in Section 2.5, rather than the simple inequality given in Equation~9}. The top panel shows a planet with a separation from a sun-like star of 0.05~AU and the bottom panel 0.1~AU. The sun-like star is assumed to have a saturated XUV flux of $10^{-3.5}~$L$_\odot$ at all ages. {\bc The white regions in the left of the plot, labelled as ``no solution (evaporation valley)'' are regions of parameter space planet where a H/He envelope would have undergone run-away loss and is a manifestation of the evaporation valley}. The dotted lines show planetary radius evolution curves (with no mass-loss) that begin at 10~R$_\oplus$ at 10~Myr and track $\tau_{\rm KH}=T_{\rm age}$.}
    \label{fig:simple_rad}
\end{figure}

In Figure~\ref{fig:simple_rad} we show the minimum mass required for the ``vanilla'' scenario where $\tau_{\rm KH}\sim T_{\rm age}$ at all ages. 
The dotted lines on these figures show the radius evolution of planets (not undergoing mass-loss), which begin at 10~R$_\oplus$ at 10~Myr. These evolutionary curves do not cross many minimum mass contours indicating that there is little strong age preference in the range of 10 to 100~Myr for selecting planets for this kind of analysis (although we will investigate this more precisely in Section~\ref{sec:mesa}). This is fairly easy to understand; as the planet cools and contracts the absorbed XUV falls, reducing the mass-loss rate. However, the total time to resist mass-loss increases. These two competing effects approximately balance, resulting in a minimum mass that does not change strongly with age. Once the XUV flux is no-longer saturated and rapidly falls with time, the mass-loss rate drops precipitously and the minimum mass will also drop rapidly with age.

The difference between the two panels in Figure~\ref{fig:simple_rad} indicates, as expected from the previous analysis, that close-in planets require higher masses. For those young planets discovered to date with radii in the range 5-10~R$_\oplus$, minimum masses in the range of 5-15~M$_\oplus$ are required.
\subsubsection{Constraining the initial Kelvin-Helmholtz timescale}

While the minimum masses provide a useful guide they do not provide much insight into planetary formation. Here, we elaborate on the much more interesting case of young planets with well measured masses and radii. 

\begin{figure}
    \centering
    \includegraphics[width=\columnwidth]{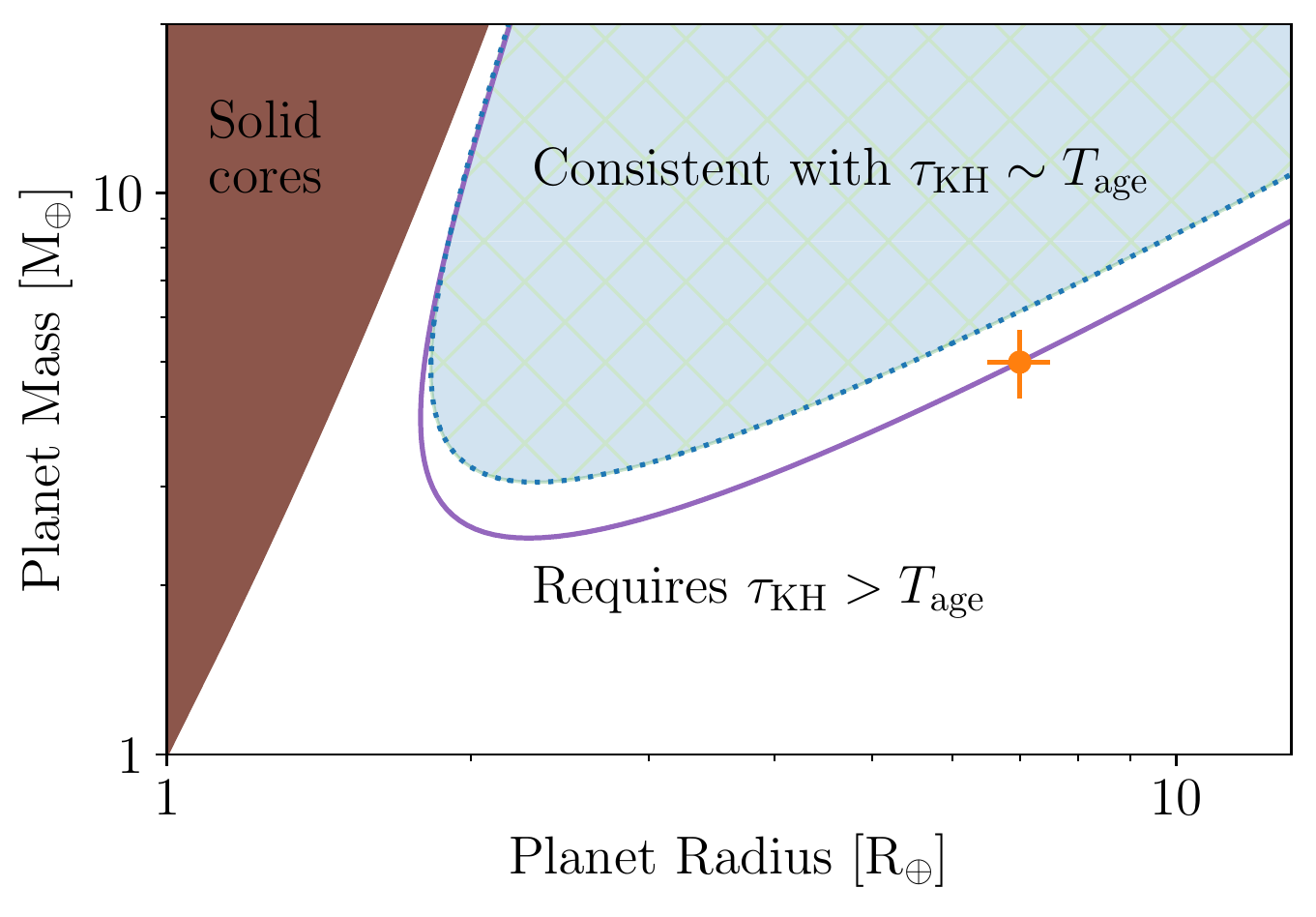}
    \caption{The planet mass and radius plane separated into those planets consistent with $\tau_{\rm KH}\sim T_{\rm age}$ and those which require longer initial Kelvin-Helmholtz timescales. {\bc As in Figure~2 this figure uses the more sophisticated calculation described in Section 2.5, rather than the simple inequality given in Equation~10 }. The diagram is shown for a planet located at 0.1~AU around an XUV saturated (10$^{-3.5}$~L$_\odot$), 50~Myr old sun-like star. Planets that sit the white region would require a long Kelvin-Helmholtz contraction timescale at formation. The point shows a representative young planet with a measured radius of 7~R$_\oplus$ and mass of 5~M$_\oplus$ shown with indicative 10\% error-bars. {\bc The solid-line and dotted line show curves with a constant $\tau_{\rm KH}$, with values of 382~Myr and 50~Myr respectively} . Thus, this representative planet would require a current (and hence formation) contraction timescale of $\gtrsim T_{\rm age}$.}
    \label{fig:tkh_simple}
\end{figure}

In Figure~\ref{fig:tkh_simple} we show how the mass-radius plane is partitioned into regions of parameter space that are consistent with $\tau_{\rm KH}\sim T_{\rm age}$ and those requiring $\tau_{\rm KH}\gtrsim T_{\rm age}$. This analysis is performed for a planet located at 0.1~AU around an XUV saturated, 50~Myr old Sun-like star. 
Placing a representative young planet with  a measured radius of 7~R$_\oplus$ and mass of 5~M$_\oplus$ on this diagram indicates it would require a longer Kelvin-Helmholtz contraction timescale (and hence lower entropy) than predicted by simple formation scenarios. Given theoretical ideas, such as ``boil-off'', predict initial Kelvin-Helmholtz contraction timescales of $\sim$ 100~Myr \citep{OW2016}, this figure indicates they could be detected with radii and mass measured to the $\sim 10$\% precision level. 

Its important to emphasise (as demonstrated in Equation~\ref{eqn:critera2}) that this type of analysis only provides a bound on the Kelvin-Helmholtz contraction timescale, where the equality holds when a planet is on the limit of stability due to mass-loss. Therefore, finding a planet in the region consistent with $\tau_{\rm KH}\sim T_{\rm age}$ does not imply that it doesn't have a longer contraction timescale, rather it could just be very stable to envelope mass-loss. 

\section{Numerical planetary evolution models}
\label{sec:mesa}
In the previous section we have used analytic tools to illustrate the basic physics; however, these models can only be pushed so far. For robust and quantitative results full numerical models are a must. This is for several reasons, most importantly, many of the transiting planets discovered to date are large and thus may contain quite significant envelope mass-fractions ($\gtrsim 10\%$). While self-gravity of such an envelope is small it is not negligible, and not included in the previous analytic model. Additionally, in the previous section we assumed an ideal equation-of-state with constant ratio of specific heats and power-law opacity model. While these choices are acceptable for understanding demographic properties, these assumptions induce unnecessary errors in the analysis of individual systems. Finally, by characterising the full evolutionary history (rather than the instantaneous state) we are able to leverage even more power. This is because not all planetary structures that are consistent with a planet's current state are consistent with its evolutionary history {\bc once mass-loss is taken into account}. The last point is demonstrated by the fact \citet{Owen2016} were able to provide a (albeit weak) constraint on the entropy of formation for the old planet {\it Kepler-}36c. {\bc Specifically in the previous section we only asked the question whether $\tau_{\rm KH}\gtrsim T_{\rm age}$. In this section, by including the full evolutionary history, we are able to compare to the planet's {\it initial} Kelvin Helmholtz timescale, which we define to be the envelope's Kelvin Helmholtz timescale at the end of disc dispersal. This comparison is more powerful, as it allows to to explore initial Kelvin-Helmholtz timescales which are shorter than the planet's current age.}

Therefore, to overcome the above shortcomings we solve for the full planetary evolution using {\sc mesa} \citep{Paxton2011,Paxton2013,Paxton2015}. The {\sc mesa} models are identical to those used in \citet{Owen2016} and \citet{OwenLai2018}, and include the impact of stellar irradiation (which tracks the \citealt{Baraffe1998} stellar evolution models) and photoevaporation using the \citet{Owen2012} mass-loss rates. 

\subsection{Example planet}
Here we return to our example planet from Figure~\ref{fig:tkh_simple}, a planet located at 0.1~AU around a 50~Myr sun-like star. Nominally, we consider this planet to have a measured radius of 7~R$_\oplus$ and measured mass of 5~M$_\oplus$, but we will investigate how changes to the mass, as well as measurement precision, will affect constraints on the planet's initial Kelvin-Helmholtz timescale.

\subsubsection{Time undergoing photoevaporation}

One of the big uncertainties at young ages is how long the planet has been exposed to XUV irradiation, and hence photoevaporating. When embedded in the protoplanetary disc it is protected from XUV photons. Thus, the age of the star only provides an upper bound on the time the planet has spent photoevaporating. Since disc lifetimes vary between $\sim 1$ and $\sim 10$~Myr, at young ages this is not a trivial uncertainty. We include this uncertainty in our analysis by deriving the probability distribution for the time a planet has spent photoevaporating after disc dispersal $T_p$. We then marginalise over this probability when determining our lower bound on the planet's initial Kelvin-Helmholtz timescale. 

We take the star to have a Gaussian age uncertainty with mean $t_*$ and error $\sigma_*$. We further assume after a time $t_d$, the disc fraction decays exponentially with the form 
\begin{equation}
    \propto \exp\left(-\frac{T_d-t_d}{\sigma_d}\right)
\end{equation}
where $T_d$ is the disc's lifetime and $\sigma_d$ is the decay time for the disc fraction. Such a phenomenological form describes the evolution of the protoplanetary disc fraction \citep[e.g.][]{Mamajek2009}. Now given a star's actual age ($T_*$) is a sum of the unknown disc's lifetime and the unknown time the planet has been undergoing photoevaporation ($T_p$), then we know $T_*=T_p+T_d$.
Therefore the probability distribution for $T_p$ can be written as:
\begin{eqnarray}
   P(T_p) &=& \frac{1}{2\sigma_d}\exp\left[\frac{\sigma_*^2+2\sigma_d\left(T_p+t_d-t_*\right)}{2\sigma_d^2}\right]\nonumber \\&\times&\left\{1-{\rm erf}\left[\frac{\sigma_*^2+\sigma_d\left(T_p+t_d-t_*\right)}{\sqrt{2}\sigma_*\sigma_d}\right]\right\}
\end{eqnarray}

In this work we set $t_d=1$~Myr and $\sigma_d=3$~Myr as this reproduces the fact that all (single) stars host discs at an age of 1~Myr, but by 10~Myr the vast majority of stars have dispersed their discs. {\bc Thus, we adopt an initial Kelvin-Helmholtz timescale of 10~Myr as the upper limit that can be reached in standard core-accretion theory.} 

\subsubsection{Results}

\begin{figure*}
\centering
\includegraphics[width=\textwidth]{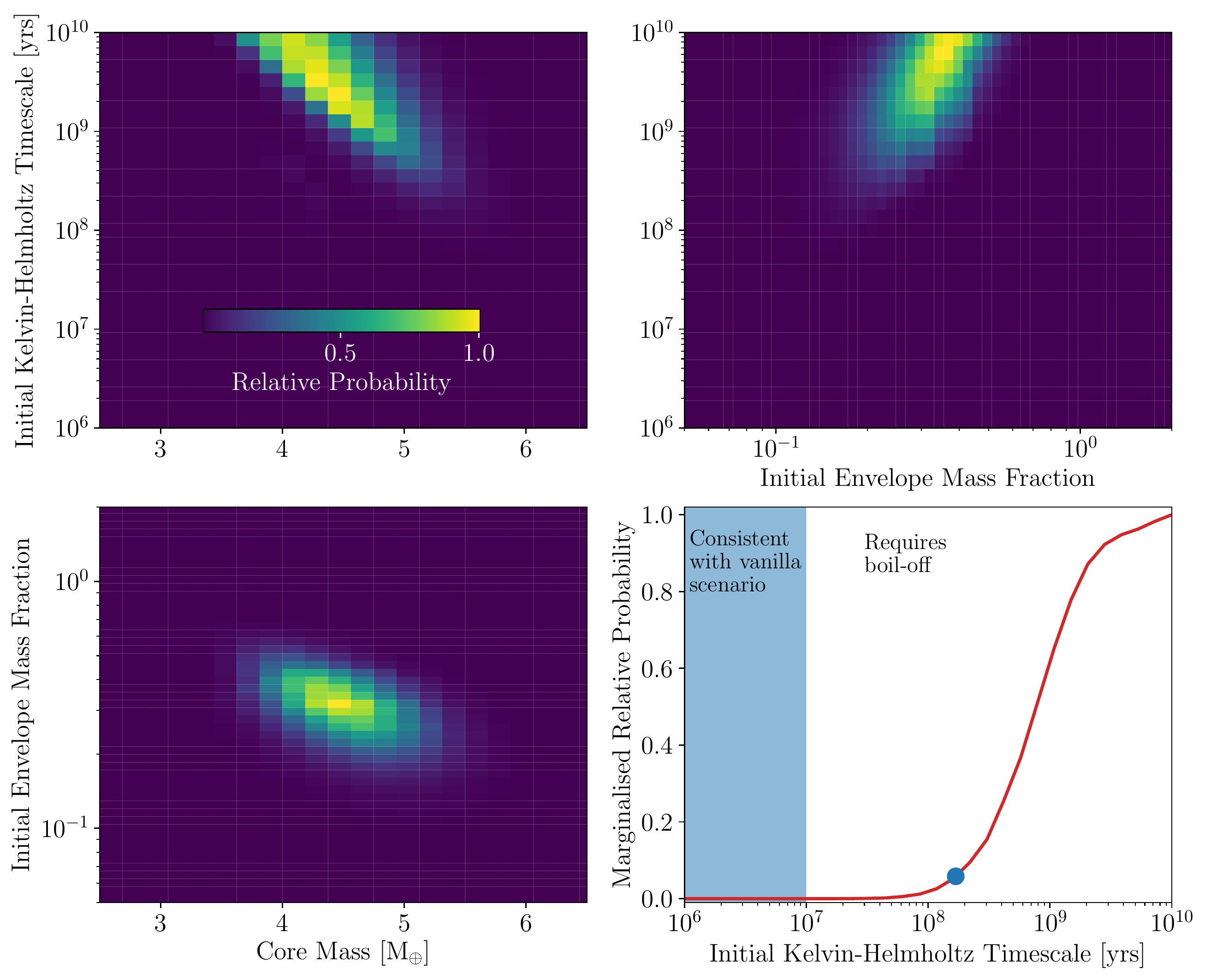}
\caption{The top left, top right and bottom left panels show joint probability distributions for the initial Kelvin-Helmholtz timescale, core mass and initial envelope mass fractions for a 50$\pm5$~Myr old planet with a radius of $7\pm0.7$~R$_\oplus$ and mass of $5\pm 0.5$~M$_\oplus$. The bottom right panel shows the marginalised probability distribution for the initial Kelvin-Helzholtz timescale, with the point indicating the 99\% lower-limit at a value of 168~Myr.  } \label{fig:big_panel}
\end{figure*}

In Figure~\ref{fig:big_panel} we show joint probability distributions for the initial Kelvin-Helmholtz timescale, initial envelope mass fraction and core mass, as well as the marginalised probability distribution for the initial Kelvin-Helmholtz timescale. This analysis has been performed assuming $10\%$ Gaussian errors on stellar age, radius and mass. Similar to our analysis in the earlier section for our 7~R$_\oplus$ and 5~M$_\oplus$ 50~Myr old planet we find that it would require an initial Kelvin-Helmholtz timescale significantly longer than would be predicted by standard core-accretion theory. In this example, we would place a 99\% lower limit on the initial Kelvin-Helmholtz timescale of $\sim 170$~Myr. The joint probability distributions are also correlated as expected with our earlier analysis. Lower mass planets require longer initial Kelvin-Helmholtz timescales and higher initial envelope mass fractions. 

\begin{figure}
    \centering
    \includegraphics[width=\columnwidth]{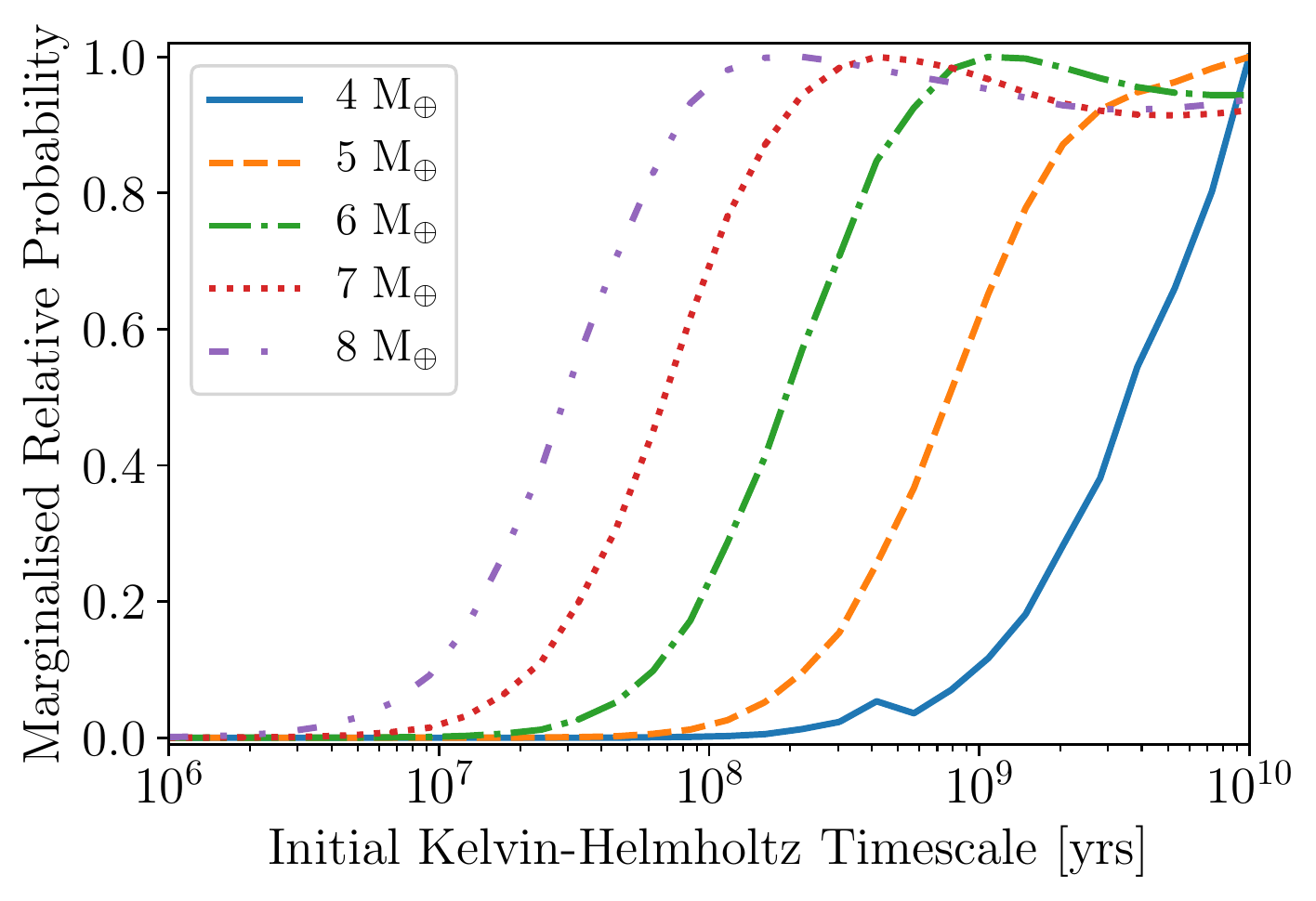}
    \caption{Marginalised probability distributions for the initial Kelvin-Helmholtz timescale for a 50$\pm5$~Myr old planet with a radius of $7\pm0.7$~R$_\oplus$. Different lines show different planet masses (with 10\% uncertainty). For this particular planet a measured mass of $\lesssim 7$~M$_\oplus$ would be required to claim evidence of boil-off.}
    \label{fig:vary_mass}
\end{figure}

We explore the role of planet mass in Figure~\ref{fig:vary_mass} where we consider measured planet masses between 4 and 8 Earth masses (again for our 7~R$_\oplus$, 50~Myr old planet with 10\% measurement uncertainties). We note very few 4~M$_\oplus$ models are consistent with the measured radius, as most have initial envelope mass fractions $\sim 1$, making them extremely vulnerable to photoevaporation \citep{Owen2019}. 
As expected, as the planet mass increases, the bound on the initial Kelvin-Helmholtz timescale decreases (as the higher-mass core is able to hold onto a less massive, and thus higher entropy envelope). 

\begin{figure}
    \centering
    \includegraphics[width=\columnwidth]{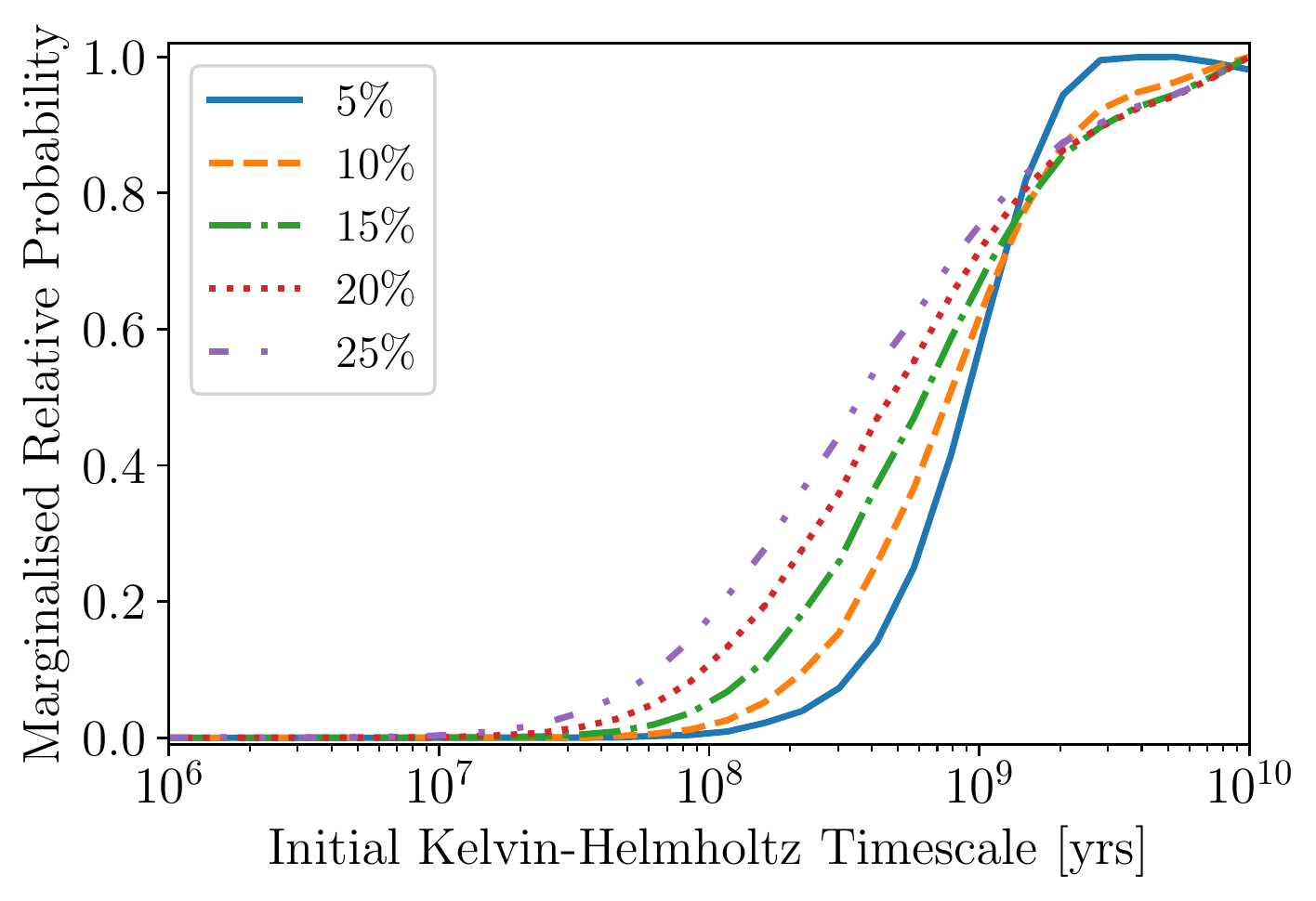}
    \caption{Marginalised probability distributions for the initial Kelvin-Helmholtz timescale for a 50$\pm5$~Myr old planet with a radius of $7\pm0.7$~R$_\oplus$ and mass of $5$~M$_\oplus$. Different lines show different uncertainties on the measured planet mass.}
    \label{fig:mass-error}
\end{figure}

While $\lesssim 10\%$ measurement uncertainties on planet radius and stellar age\footnote{Much of the uncertainty in the time a planet has spent photoevaporating is dominated by the uncertainty in the disc dispersal timescale at ages $\lesssim 100$~Myr.} have been achieved for known young planets, stellar activity may mean obtaining radial-velocity mass measurements at a $\sim 10$\% precision is difficult. Therefore, in Figure~\ref{fig:mass-error} we show how sensitive our constraints on the initial Kelvin-Helmholtz timescale are to mass uncertainties in the range of 5-25\%. As you would naturally expect, increasing the uncertainty means higher mass planets become consistent with the measured mass, allowing shorter initial Kelvin-Helmholtz timescales. However, even with a tentative $\sim 25\%$ mass detection, for this example we would still be able to place a useful constraint on the initial Kelvin-Helmholtz timescale. 

This gives us confidence that measured masses can place useful constraints on the entropy of formation of young transiting planets, even if those mass measurements are tentative. 

\subsection{What age is best?}

One question that remains is what is the best system age to do this experiment for? Obviously young planets allow you to constrain shorter and shorter initial Kelvin-Helmholtz timescales as they've had less chance too cool. Yet, at young ages there are two confounding effects. First photoevaporation may not have had enough time to significantly control the planet's evolution. Second, at very young ages, the time the planet has spent photoevaporating after disc dispersal is not dominated by the uncertainty in the age of the system, but rather by the unknown disc lifetime. For example a 10~Myr old planet could have spent anywhere between 0 and $\sim 9$~Myr photoevaporating. However, wait too long and the planet will have cooled sufficiently that knowledge of its initial thermodynamic state will have been lost, especially at ages $\gtrsim 100$~Myr when photoevaporation no-longer dominates. 

\begin{figure}
    \centering
    \includegraphics[width=\columnwidth]{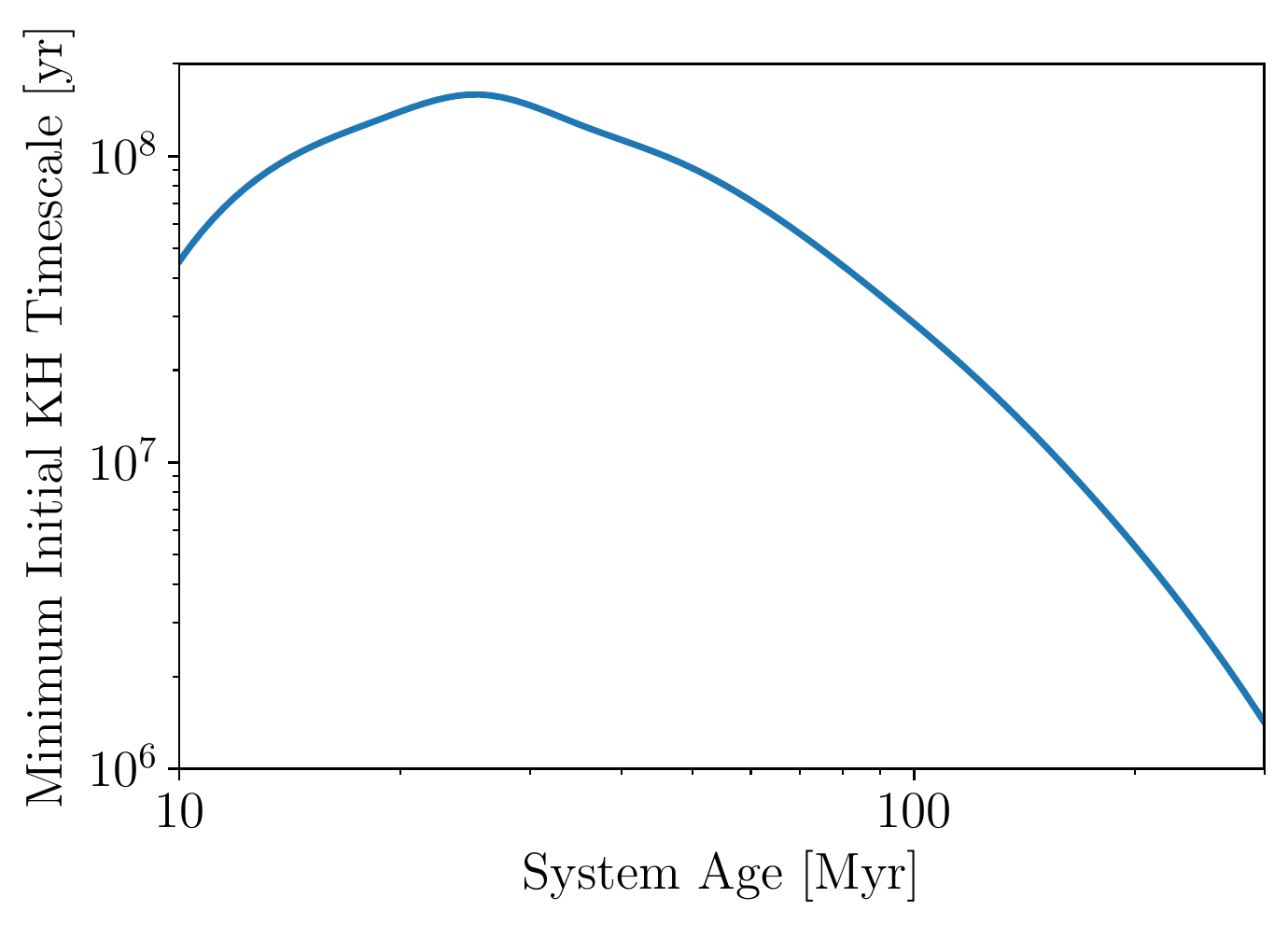}
    \caption{The constraint on the minimum inferred initial Kelvin-Helmholtz timescale as a function of stellar age. }
    \label{fig:best_age}
\end{figure}

Thus, we expect there to be an optimum range of stellar ages at which this experiment is most stringent. In order to assess this we take the evolution of a planet with a 4.375~M$_\oplus$ core, with an initial envelope mass fraction and Kelvin-Helmholtz timescale of $0.3$ and $500$~Myr respectively. This model roughly corresponds to our 5~M$_\oplus$, 7~R$_\oplus$, 50 Myr old planet studied earlier. We then use our method to constrain its initial Kelvin-Helmholtz timescale as a function of age assuming 10\% errors on planet mass, radius and stellar age. The minimum initial Kelvin-Helmholtz timescale (at the 99\% confidence level) is shown as a function of age for this exercise in Figure~\ref{fig:best_age}.
The system age that is most constraining lies around $\sim 30$~Myr, where even with errors and a fairly robust confidence level we recover the actual initial Kelvin-Helmholtz timescale within a factor of 3.  Ages in the range $\sim 20-60$~Myr have a constraint that varies by less than $50\%$ of the absolute maximum. We can clearly see that after 100~Myr the constraint on the initial Kelvin-Helmholtz timescale becomes uninformative. Therefore, for real systems our method should provide meaningful constraints on the initial Kelvin-Helmholtz timescale and hence formation entropy for planets around stars with ages in the range 20-60~Myr.

\section{Application to real planets}
Having shown that by using the photoevaporation model it is possible to constrain a young planet's entropy of formation we turn our attention to detected young planets, and consider how their inferred entropy of formation varies as a function of possible measured mass. We choose to focus here on DS Tuc Ab and V1298 Tau c, out of the handful of known young planets, as these are the most strongly irradiated, and therefore most likely to result in strong constraints on their initial Kelvin-Helmholtz contraction timescale.

\subsection{DS Tuc Ab}

\begin{figure}
    \centering
    \includegraphics[width=\columnwidth]{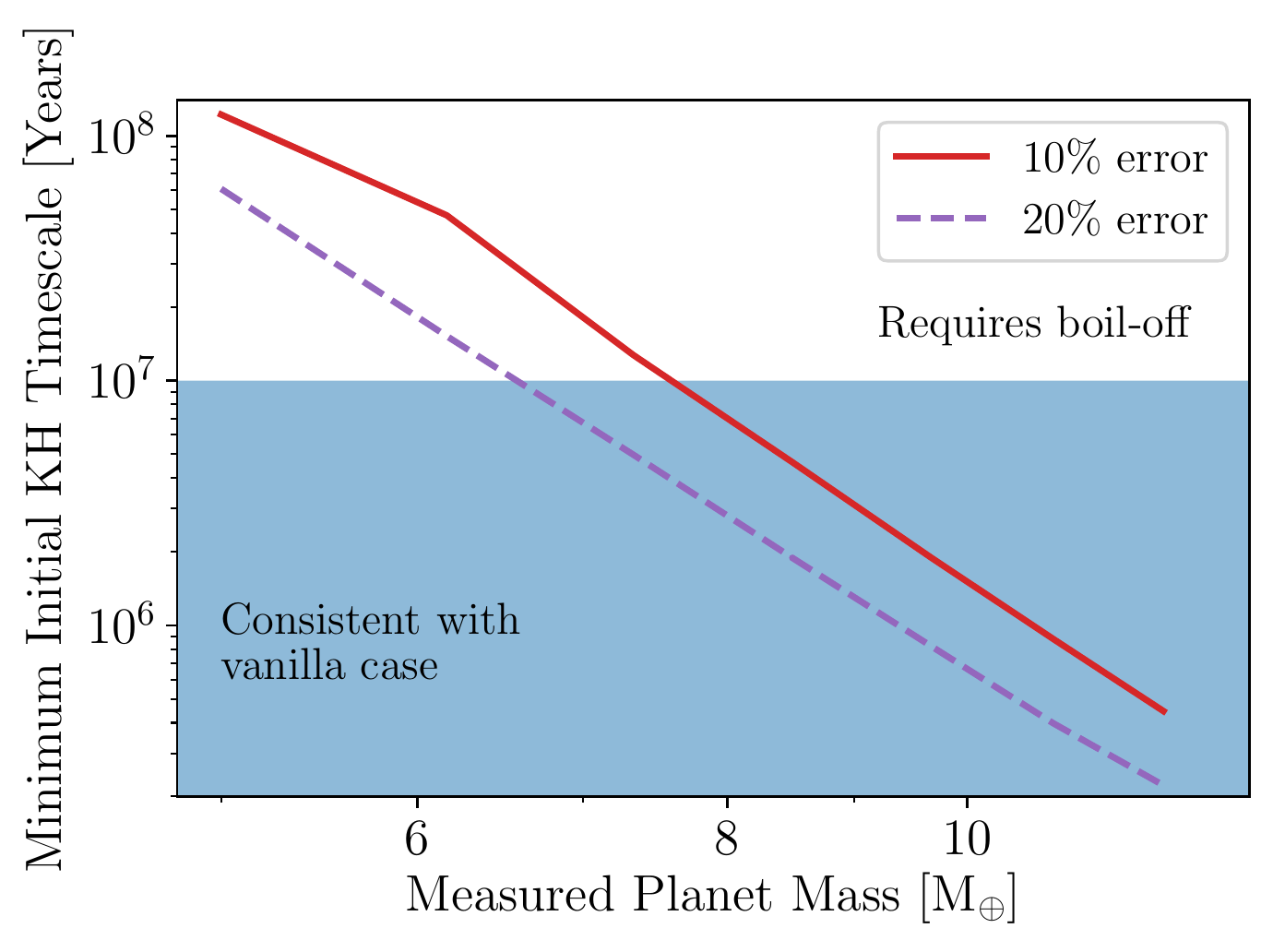}
    \caption{The minimum initial Kelvin-Helmholtz timescale (at the 99\% confidence limit) for DS Tuc Ab shown as a function of measured planet mass for a 10\% and 20\% measured mass uncertainty. A measured mass of $\lesssim 7-8$M$_\oplus$ would require a ``boil-off'' phase to explain.  }
    \label{fig:limit_DS}
\end{figure}

DS Tuc Ab \citep{Benatti2019,Newton2019} is a $5.70\pm0.17$~R$_\oplus$ planet\footnote{We choose to use the stellar and planetary parameters from \citet{Newton2019}.} discovered around a $45\pm4$~Myr, 1.01~M$_\odot$ star, orbiting with a period of 8.1~days. Using exactly the same formalism as applied in Section~\ref{sec:mesa} we consider the constraints on entropy of formation and initial Kelvin-Helmholtz timescale as a function of planet mass. 
We find that a measured mass $\lesssim 4.5$~M$_\oplus$ would be inconsistent with the current properties of DS Tuc Ab. In Figure~\ref{fig:limit_DS} we show how the inferred lower-limit on the initial Kelvin-Helholtz timescale varies with both the measured planet mass and the measurement uncertainty. Our results indicate that a measured mass $\lesssim 7.5$~M$_\oplus$ with a uncertainty of 10\% (or $\lesssim$ 6.5~M$_\oplus$ with a 20\% uncertainty) would require a longer than na\"ively expected initial Kelvin-Helmholtz timescale and require with a ``boil-off'' phase. A mass of 7.5~M$_\oplus$ corresponds to a radial velocity semi-amplitude of $\sim 2.4$~m~s$^{-1}$, eminently detectable with current instrumentation, stellar noise not withstanding.  

\subsection{V1298 Tau c}

\begin{figure}
    \centering
    \includegraphics[width=\columnwidth]{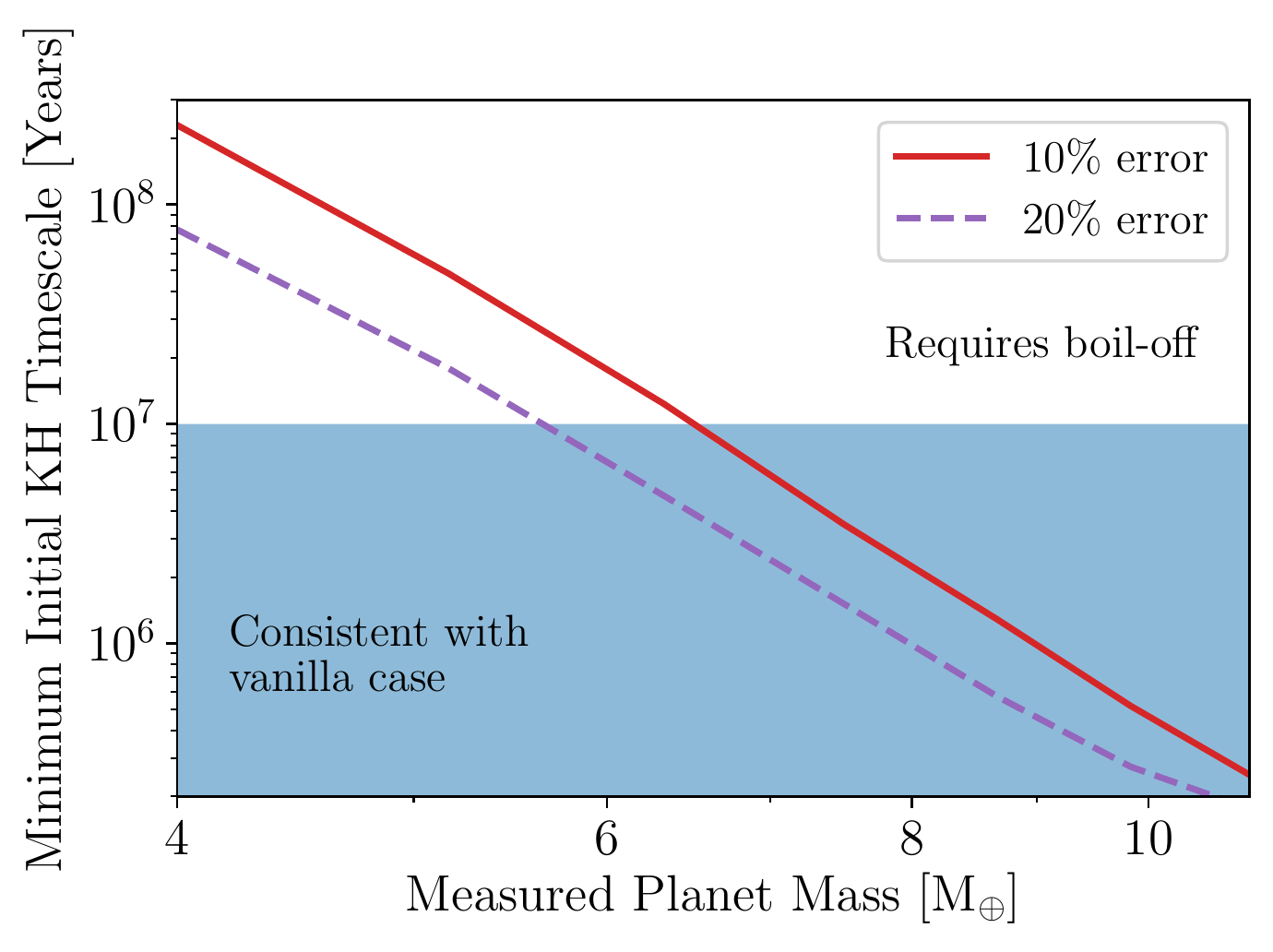}
    \caption{The minimum initial Kelvin-Helmholtz timescale (at the 99\% confidence limit) for V1298 Tau c shown as a function of measured planet mass for a 10\% and 20\% measured mass uncertainty. A measured mass of $\lesssim 6-7$M$_\oplus$ would require a ``boil-off'' phase to explain.  }
    \label{fig:V1298_result}
\end{figure}

The 23$\pm4$~Myr old, 1.1~M$_\odot$, V1298 Tau system contains four large transiting young planets \citep{David2019}. All planets are between 5-11~R$_\oplus$ in radii and orbit close to the star with periods $\lesssim 100$~days indicating it is likely to be a precursor to the the archetypal {\it Kepler} multi-planet systems. Given our analysis in Section~\ref{sec:min_tkh} indicated that planets much closer to their star will provide the most stringent limits (due to more vigorous photoevaporation) we select planet c to investigate here. V1298 Tau c is a $5.59\pm0.34$~R$_\oplus$ planet with an orbital period of 8.2~days. Since the V1298 Tau system is a multi-planet system, dynamical arguments have already put constraints on the sum of planet c's and d's mass to be $7^{+21}_{-5}$~M$_\oplus$. Like DS Tuc Ab above we calculate the minimum initial Kelvin-Helmholtz timescale, taken to be the 99\% lower limit, as a function of measured planet mass (with both 10\% and 20\% measurement uncertainties) which is shown in Figure~\ref{fig:V1298_result}

We note measured planet masses $\lesssim 4$~M$_\oplus$ are inconsistent with V1298 Tau c's current properties. A mass measurement of $\lesssim 6.5$~M$_\oplus$  with a 10\% uncertainty (or $\lesssim$5.5~M$_\oplus$ with a 20\% uncertainty) would require a boil-off phase to explain. This corresponds to a RV semi-amplitude of $\sim 2$~m~s$^{-1}$, again within the realm of possibility for radial velocity characterisation (stellar noise not withstanding). Further, since V1298 Tau is a multi-planet system, this permits the possibility of mass constraints through Transit Timing Variations. 

\section{Discussion}

Typical sub-neptune and super-earth planets are expected to be much larger at early ages.
In this work we have shown that with a combined mass and radius measurement of a  proto-sub-neptune/super-earth ($\lesssim 100$~Myr old), a lower bound can be placed on its initial Kelvin-Helmholtz timescale. This lower bound provides valuable insight into the accretion and early evolution of its H/He envelope. This lower bound is essentially found by answering how low-mass a H/He envelope can exist on the planet given it's been undergoing photoevaporation. For a fixed planet radius, a higher entropy envelope contains less mass and is therefore more vulnerable to mass-loss. Whereas, lower entropy envelopes need to be more massive and thus are able to resist mass-loss for longer. 

One might expect that the younger the planet this experiment is done for the tighter a constraint can be obtained. While this is generally true, at the youngest ages the fact protoplanetary disc lifetimes vary means one cannot be certain how long a planet has been undergoing mass-loss. Thus, we find that the optimum ages for this experiment are around 20-60~Myr. 

Further, more accurate measurements obviously result in tighter constraints. In this work, we showed that measurement precision in the range of 10-20\% on radius, mass and age are required to perform this analysis. Current (and previous) transit surveys have already reached and exceeded this requirement on known young planets (e.g. the planets in the V1298 Tau system have radius uncertainties in the range 6-7\%). Further, the already published young planets have age uncertainties at the 10\% level. 

What is difficult to ascertain is whether the mass measurements at the $\lesssim 20\%$ precision are achievable. As discussed in Section~4, the problem is not RV precision. Rather it is intrinsic stellar variability, particularly due to spots \citep[e.g.][]{Huerta2008}, which are more prevalent on younger stars. Recent work using Gaussian Processes have shown that it is possible to model intrinsic stellar variability and obtain mass measurements for planets \citep{Haywood2014,Grunblatt2015}. Using this technique \citet{Barragan2019} recently obtained an RV mass measurement for K2-100b, which is a moderately young star showing significant intrinsic RV variability. Alternatively, if the planets happen to reside in multi-transiting systems, then TTVs could be used. While we acknowledge the difficulty in obtaining the mass measurements we require, we advocate that the scientific value of constraints on planets' initial entropy is important enough to motivate the effort. 

Since we are using photoevaporation to constrain the entropy of formation, our results are sensitive to the accuracy of theoretical photoevaporation calculations. In this work we use the mass-loss rates of \citet{Owen2012} which are consistent with the location and slope of the ``evaporation valley'' \citep{VanEylen2018} \footnote{\bc As does the core-powered mass-loss model \citep{Gupta2019,Gupta2019b}.}, and are generally in good agreement with observed outflows \citep{Owen2012}. Only more theoretical and observational work calibrating photoevaporation models  can assess the impact changing the mass-loss rates may have on entropy constraints.

\subsection{Links to planet formation theory}

Since the discovery of sub-neptunes and super-earths there has been much work on their origin \citep[e.g.][]{Ikoma2006,Ikoma2012,Lee2014,Venturini2015,Venturini2016,Ginzburg2018}. It is clear that the only way to explain (at least some of) their current densities is to have a have a planetary core (made of some mixture of rock, iron and ices) surrounded by a H/He envelope which contains $\sim 1-10$\% of the planet's total mass \citep[e.g.][]{JontofHutter2016}. 

Such a planetary composition would naturally arise through the core-accretion mechanism, whereby the growing solid core accretes a H/He envelope over the disc's lifetime. In this standard picture, the accreting planetary envelope smoothly connects to the disc, but remains in quasi-hydrostatic and thermal equilibrium. As the envelope cools, it contracts and slowly accretes. This process happens on the envelope's  Kelvin-Helmholtz timescale, which without any strong internal heating sources, quickly equilibrates to roughly the envelope's age. If disc dispersal allows the envelope to remain in quasi-hydrostatic and thermal equilibrium, then a planet's ``initial Kelvin-Helmholtz timescale'' (which we define as the Kelvin-Helmholtz timescale after disc dispersal) is essentially the time it has spent forming, which is bounded by the protoplanetary disc lifetime (e.g. $\lesssim 10$~Myr). 

While the basic picture appears to fit, there is growing evidence that the standard core accretion model significantly over-predicts the amount of H/He a core of a given mass should accrete \citep{Jankovic2019,Ogihara2020,Alessi2020,Rogers2020}. In some cases the problem is so acute that it's not clear why certain planets did not become giant planets \citep[e.g.][]{Lee2014,Lee2019}. Several solutions have been proposed to solve this problem.  \cite{Chen2020} suggested enhanced opacity from dust could slow the atmosphere's accretion.  Using numerical simulations, \citet{Ormel2015} and \citet{Fung2015} suggested that the envelope was not in quasi-hydrostatic equilibrium with the disc, but rather high-entropy disc material continually flowed into the envelope, preventing it from cooling. 

\cite{Lee2016} hypothesised instead these planets do not spend the entire disc lifetime accreting from the nebula, but rather formed rapidly (over a timescale of $10^5-10^6$~years) in the final ``transition'' disc stage of the protoplanetary disc. The much lower gas surface densities and the shorter lifetime of the transition disc phase gave rise to smaller accreted atmospheres. The above modifications to the ``vanilla'' core accretion theory model will typically result in higher entropy envelopes and therefore initial Kelvin-Helmholtz contraction timescales, significantly shorter than the standard value of a few Myr.

An alternative solution to the over accretion problem\footnote{Although it does not prohibit the modifications to core-accretion theory described above.} is the introduction of additional mass-loss. While it does not seem energetically feasible to increase the rates of either photoevaporation or core-powered mass-loss (as they are already fairly efficient), the assumption that the envelope maintains some sort of dynamical equilibrium as the disc disperses seems unlikely. Protoplanetary discs are observed to live and evolve slowly over their 1-10~Myr lifetimes. However, the dispersal process is rapid with a timescale of $\sim 10^5$~years \citep[e.g.][]{kenyon95,ercolano11,koepferl13,owenreview2016}. 

As argued by \citet{OW2016} and \citet{Ginzburg2016} this means accreted H/He envelopes cannot maintain dynamical and thermal balance with the gas in the dispersing disc. As such the envelopes become over-pressurised, and expand hydrodynamically into the forming cicumstellar vacuum. This ``boil-off'' process results in mass-loss (in extreme cases up to 90\% of the initial envelope is lost), but also importantly cooling of the interior. This is because the bottleneck for cooling (the radiative-convective boundary) is replaced by an advective-convective boundary and thermal energy is removed from the interior quickly by advection and mass-loss. Using simulations, \citet{OW2016} found that after this boil-off process, the remaining envelopes had their entropies reduced. Their Kelvin-Helmholtz contraction timescales at the end of disc dispersal were around $\sim 100$~Myr. 

Thus, any constraints of the initial Kelvin-Helmholtz contraction timescale of proto-sub-neptunes/super-earths will be invaluable for constraining and testing our models for their origins.  

\section{Summary}

The formation of sub-Neptunes and super-Earths is uncertain and many formation models have been proposed to explain their origin. These formation models are essentially unconstrained by the old, evolved exoplanet population that has a typical age of 3~Gyr. However, various planet formation models predict vastly different entropies at the end of protoplanetary disc dispersal. Characterising the entropies at the end of disc dispersal in terms of initial Kelvin-Helmholtz contraction timescales, these predictions range from $\lesssim 1$~Myr to $\gtrsim 100$~Myr.

A young proto-sub-neptune/super-earth with a measured mass, radius and age can be used to place a lower bound on its initial Kelvin-Helmholtz contraction timescale. This requires the planet to be close enough to its host star that photoevaporation has had an impact on its evolution. This constraint is obtained by answering how low-mass a H/He envelope can exist on the planet given the mass-loss it experienced. For a fixed planet radius, a higher entropy envelope contains less mass and is therefore more vulnerable to mass-loss. Whereas, lower entropy envelopes need to be more massive and thus are able to resit mass-loss for longer.

We have shown that planets around host stars with ages 20-60~Myr are the optimum targets for this kind of analysis. Applying our hypothesised method to detected young planets DS Tuc Ab and V1298 Tau c we show planet mass constraints (with $\lesssim 20\%$ precision) in the range 7-10~M$_\oplus$ would be consistent with our standard picture of core-accretion. Mass measurements $\lesssim 7$~M$_\oplus$ would favour a ``boil-off'' process, where a planet loses mass and its interior cools significantly during dispersal. 

While precise mass measurements of low-mass planets orbiting young stars are likely to be challenging, the insights into planet formation that could be obtained warrant the effort.

\section*{Acknowledgements}
JEO is supported by a Royal Society University Research Fellowship and a 2019 ERC starting grant (PEVAP).

\section*{Data Availability}
The code used to create the planet structure models in Section~2.5 is freely available at: \url{https://github.com/jo276/EvapMass}. The custom {\sc mesa} code used to calculate the planet evolution models in Section~3 and 4 is freely available at: \url{https://github.com/jo276/MESAplanet}.
The remaining data underlying this article will be shared on reasonable request to the corresponding author.




\bibliographystyle{mnras}
\bibliography{bib_paper}



\appendix



\bsp	
\label{lastpage}
\end{document}